\documentclass[aps,preprint,nofootinbib,floatfix,superscriptaddress]{revtex4}
\usepackage{epsfig}
\usepackage{graphicx}
\usepackage{multirow}
\usepackage{latexsym,amsbsy,amsmath}
\usepackage{stackrel}
\begin{document}
\title{Charmed baryon$-$nucleon interaction}
\author{H.~Garcilazo} 
\email{humberto@esfm.ipn.mx} 
\affiliation{Escuela Superior de F\' \i sica y Matem\'aticas, \\ 
Instituto Polit\'ecnico Nacional, Edificio 9, 
07738 Mexico D.F., Mexico} 
\author{A.~Valcarce} 
\email{valcarce@usal.es} 
\affiliation{Departamento de F\'\i sica Fundamental and IUFFyM,\\ 
Universidad de Salamanca, E-37008 Salamanca, Spain}
\author{T.~F.~Caram\'es} 
\email{carames@usal.es} 
\affiliation{Departamento de F\'\i sica Fundamental and IUFFyM,\\ 
Universidad de Salamanca, E-37008 Salamanca, Spain}
\date{\today} 
\begin{abstract}
We present a comparative study of the charmed baryon$-$nucleon interaction 
based on different theoretical approaches.
For this purpose, we make use of i) a constituent quark model tuned in 
the light-flavor baryon$-$baryon interaction and the hadron spectra, 
ii) existing results in the literature based both on 
hadronic and quark-level descriptions,
iii) (2+1)-flavor lattice QCD results
of the HAL QCD Collaboration at unphysical pion masses 
and their effective field theory extrapolation to the physical pion mass.
There is a general qualitative agreement among the different 
available approaches to the charmed baryon$-$nucleon interaction. Different from 
hadronic models based on one-boson exchange potentials, quark$-$model based results 
point to soft interactions without two-body bound states. They
also support a negligible channel coupling, due either to tensor forces
or to transitions between different physical channels, $\Lambda_c N - \Sigma_c N$.
Short-range gluon and quark-exchange dynamics
generate a slightly larger repulsion in the $^1S_0$ than in the $^3S_1$ $\Lambda_c N$ 
partial wave. A similar asymmetry 
between the attraction in the two $S$ waves of the $\Lambda_c N$ interaction also appears 
in hadronic approaches.
A comparative detailed study of Pauli suppressed partial waves, as the $^1S_0 (I=1/2)$ 
and $^3S_1 (I=3/2)$ $\Sigma_c N$ channels, would help to disentangle the short-range 
dynamics of two-baryon systems containing heavy flavors.
The possible existence of charmed hypernuclei is discussed.
\end{abstract}
\maketitle 
\section{Introduction}
\label{secI}

There has been an impressive experimental progress in the spectroscopy of heavy 
hadrons, mainly in the charm sector. The theoretical analysis
of hidden and open heavy flavor hadrons has revealed how interesting is the 
interaction of heavy hadrons, with presumably a long-range part of Yukawa type, and a 
short-range part mediated by quark$-$quark and quark$-$antiquark forces.
Some of the recently reported states might appear as bound states 
or resonances in the scattering of two hadrons with heavy flavor content.
See Refs.~\cite{Liu19,Bri16,Ric16,Ric17,Ali17,Esp17} for recent overviews and discussions.
Thus, the understanding of the baryon$-$baryon interaction in the heavy flavor
sector is a key ingredient in our quest to describing the properties of hadronic matter. 

The research programs at various facilities are expected to improve our knowledge on 
the hadron$-$hadron interactions involving heavy flavors, particularly in the charm sector. 
Thus, the LHCb Collaboration at the Large Hadron Collider (LHC)
is engaged in an extensive program aimed at the analysis of charmed hadrons 
produced in the environment of high-energy proton$-$proton collisions~\cite{Ogi15}. 
The observation of five new narrow excited $\Omega_c$ 
states has already been reported~\cite{Aai17}, some of which are suggested as 
molecules containing a charmed hadron~\cite{Liu19,Bri16,Ric16,Ric17,Ali17,Esp17}. 
The planned installation of a 50~GeV high-intensity proton beam at Japan Proton Accelerator Research 
Complex (J-PARC)~\cite{Nou17,Fuj17} intends to produce charmed hypernuclei, in which 
a $Y_c$ baryon ($\Lambda_c$ or $\Sigma_c$) is bound to a nucleus. There are also planned experiments by the 
$\overline{\rm P}$ANDA Collaboration at the Facility for 
Antiproton Ion Research (FAIR)~\cite{Wie11,Hoh11} to produce charmed hadrons by annihilating 
antiprotons on nuclei. 

In addition to the recent interest in the hadron$-$hadron interaction involving heavy flavors, 
there is a long history of speculations as regards bound nuclear systems with a charmed 
baryon. The observation of events that could be interpreted
in terms of the decay of a charmed nucleus~\cite{Tip75,Bat81}, fostered conjectures
about the  possible existence of charm analogs of strange 
hypernuclei~\cite{Dov77,Iwa77,Gat78}. 
This resulted in several theoretical estimates about the binding
energy and the potential-well depth of charmed 
hypernuclei based on one-boson exchange potentials for the
charmed baryon$-$nucleon interaction~\cite{Bha81,Ban82,Ban83,Gib83,Sta86}.
The current experimental prospects have reinvigorated studies of the low-energy 
$Y_c N$ interactions~\cite{Cai03,Tsu04,Kop07,Liu12,Oka13,Hua13,Gar15,Mae16,Mae18,Vid19}.
See also the recent reviews~\cite{Hos17,Kre18}.

As pointed out by Bjorken~\cite{Bjo85} one should strive to study
systems with heavy flavors because due to their size the
quark$-$gluon coupling constant is small and therefore the
leading term in the perturbative expansion is enough to
describe the system. However, our ability of making first-principles 
analytical calculations of nonperturbative 
QCD phenomena is very limited. 
When combined with the lack of experimental information 
on the elementary $Y_c N$ interactions
there is room for some degree of speculation in the study of processes involving charmed hadrons. 
Thus, the situation can be ameliorated with the use of well constrained models based as much 
as possible on symmetry principles and analogies with other similar processes, which 
is still a valid alternative for making progress.

Within such a perspective, 
in this work we present the first comparative study of the charmed baryon$-$nucleon
interaction based on different theoretical approaches.
We employ a widely used constituent quark model (CQM)~\cite{Val05,Vij05} 
providing a good description of the low-lying spectrum of light and charmed 
hadrons~\cite{Vac05,Val08} as well as the nucleon-nucleon interaction~\cite{Gar99,Val05}.
In addition, we consider different scattered results available in
the literature. In particular, we compare to the hadronic description based 
on one-boson exchange potentials of Ref.~\cite{Liu12}; the quark-level approach relying on the
quark delocalization color screening model (QDCSM) of Ref.~\cite{Hua13};
the hybrid model of Ref.~\cite{Mae16} based on one-boson exchange potentials supplemented 
by a global short-range repulsion of quark origin; and the recent charmed baryon$-$nucleon
potential based on a $SU(4)$ extension of the meson-exchange hyperon-nucleon 
potential {\em \~A} of the J\"ulich group~\cite{Reu94} of Ref.~\cite{Vid19}. We will also consider
the recent lattice QCD simulations of the $Y_c N$ interactions 
by the HAL QCD Collaboration~\cite{Miy15,Miy16,Miy18,Mia18}.
However, the lattice QCD simulations are still obtained with unphysical pion masses.  
They have been extrapolated to the physical pion mass
using a chiral effective field theory (EFT)~\cite{Hai18}.

The paper is organized as follows. In Sect.~\ref{secII} we outline the basic ingredients 
of the CQM used to derive the $Y_c N$ interactions. We also describe the integral equations of the coupled 
$\Lambda_c N - \Sigma_c N$ system. In Sect.~\ref{secIII} we present and discuss 
the results for the $\Lambda_c N$ and $\Sigma_c N$ interactions. We show the results 
of the CQM in comparison to the available results from other theoretical approaches 
in the literature. We analyze the consequences of the different 
approaches for the possible existence of charmed hypernuclei. Finally, 
in Sect.~\ref{secIV} we summarize the main conclusions of our work.

\section{Formalism}
\label{secII}

\subsection{The quark$-$quark interaction}
\label{secIIa}

The two-body $Y_c N$ interactions are obtained from the chiral constituent quark model of Ref.~\cite{Val05}.
The model was proposed in the early 1990s in an attempt to obtain a simultaneous description 
of the light baryon spectrum and the nucleon-nucleon interaction. It was later on generalized 
to all flavor sectors~\cite{Vij05}. In this model, hadrons are described as clusters of three 
interacting  massive (constituent) quarks. The masses of the quarks are generated by the 
dynamical breaking of the original $\mathrm{SU}(2)_{L}\otimes \mathrm{SU}(2)_{R}$ chiral symmetry of the QCD 
Lagrangian at a momentum scale of the order of $\Lambda_{\rm CSB} = 4\pi f_\pi \sim 1$~GeV, 
where $f_\pi$ is the pion electroweak decay constant. For momenta typically below that 
scale, when using the linear realization of chiral symmetry, light quarks interact through 
potentials generated by the exchange of pseudoscalar Goldstone 
bosons ($\pi$) and their chiral partner ($\sigma$): 
\begin{equation}
V_{\chi}(\vec{r}_{ij})\, = \, V_{\sigma}(\vec{r}_{ij}) \, + \, V_{\pi}(\vec{r}_{ij}) \, ,
\end{equation}
where
\begin{equation}
V_{\sigma}(\vec{r}_{ij}) =
    -\dfrac{g^2_{\rm ch}}{{4 \pi}} \,
     \dfrac{\Lambda^2}{\Lambda^2 - m_{\sigma}^2}
     \, m_{\sigma} \, \left[ Y (m_{\sigma} \,
r_{ij})-
     \dfrac{\Lambda}{{m_{\sigma}}} \,
     Y (\Lambda \, r_{ij}) \right] \,, \nonumber
\end{equation}
\begin{eqnarray}
V_{\pi}(\vec{r}_{ij})&=&
     \dfrac{ g_{\rm ch}^2}{4
\pi}\dfrac{m_{\pi}^2}{12 m_i m_j}
     \dfrac{\Lambda^2}{\Lambda^2 - m_{\pi}^2}
m_{\pi}
     \Biggr\{\left[ Y(m_{\pi} \,r_{ij})
     -\dfrac{\Lambda^3}{m_{\pi}^3} Y(\Lambda
\,r_{ij})\right]
     \vec{\sigma}_i \cdot \vec{\sigma}_j 
\nonumber \\
&&   \qquad\qquad +\left[H (m_{\pi} \,r_{ij})
     -\dfrac{\Lambda^3}{m_{\pi}^3} H(\Lambda
\,r_{ij}) \right] S_{ij}
     \Biggr\}  (\vec{\tau}_i \cdot \vec{\tau}_j)
\, .
\end{eqnarray}
$g^2_{\rm ch}/4\pi$ is the chiral coupling constant, $m_i$ are the
masses of the constituent quarks, $\Lambda \sim \Lambda_{\rm CSB}$,  
$Y(x)$ is the standard Yukawa function defined by $Y(x)=e^{-x}/x$, 
$H(x)=(1+3/x+3/x^2)\,Y(x)$, and $S_{ij} \, = \, 3 \, ({\vec \sigma}_i \cdot
{\hat r}_{ij}) ({\vec \sigma}_j \cdot  {\hat r}_{ij})
\, - \, {\vec \sigma}_i \cdot {\vec \sigma}_j$ is
the quark tensor operator.

Perturbative QCD effects are taken into account through the one-gluon-exchange (OGE) 
potential~\cite{Ruj75}:
\begin{equation}
V_{\rm OGE}({\vec{r}}_{ij}) =
        {\frac{\alpha_\mathrm{s}}{4}}\,{\vec{\lambda}}_{i}^{c} \cdot {\vec{\lambda}}_{j}^{c}
        \Biggl[ \frac{1}{r_{ij}}
        - \dfrac{1} {4} \left(
{\frac{1}{{2\,m_{i}^{2}}}}\, +
{\frac{1}{{2\,m_{j}^{2}}}}\,
        + {\frac{2 \vec \sigma_i \cdot \vec
\sigma_j}{3 m_i m_j}} \right)\,\,
          {\frac{{e^{-r_{ij}/r_{0}}}}
{{r_{0}^{2}\,\,r_{ij}}}}
        - \dfrac{3 S_{ij}}{4 m_i m_j r_{ij}^3}
        \Biggr] \, ,
\label{OGE}
\end{equation}
where $\vec\lambda^{c}$ are the $\mathrm{SU}(3)$ color matrices, 
$r_0=\hat r_0/\nu$ is a flavor-dependent regularization scaling with the 
reduced mass $\nu$ of the interacting pair, and $\alpha_s$ is the
scale-dependent strong coupling constant given by~\cite{Vij05},
\begin{equation}
\alpha_s(\nu)={\frac{\alpha_0}{\rm{ln}\left[{({\nu^2+\mu^2_0})/
\gamma_0^2}\right]}} \, ,
\label{asf}
\end{equation}
where $\alpha_0=2.118$, 
$\mu_0=36.976$ MeV and $\gamma_0=0.113$ fm$^{-1}$. This equation 
gives rise to $\alpha_\mathrm{s}\sim0.54$ for the light-quark sector,
$\alpha_\mathrm{s}\sim0.43$ for $uc$ pairs, and
$\alpha_\mathrm{s}\sim0.29$ for $cc$ pairs.

Finally, any model imitating QCD should incorporate
confinement. Although it is a very important term from the spectroscopic point of view,
it is negligible for the hadron$-$hadron interaction. Lattice QCD calculations 
suggest a screening effect on the potential when increasing the interquark 
distance~\cite{Bal01} which is modeled here by, 
\begin{equation}
V_{\rm CON}(\vec{r}_{ij})= -a_{c}\,(1-e^{-\mu_c\,r_{ij}})\,
(\vec{\lambda^c}_{i}\cdot \vec{ \lambda^c}_{j})\,,
\end{equation}
where $a_{c}$ and $\mu_c$ are the strength and range parameters.
Once perturbative (one-gluon exchange) and nonperturbative (confinement and
dynamical chiral symmetry breaking) aspects of QCD have been incorporated, 
one ends up with a quark$-$quark interaction of the form,
\begin{equation} 
V_{q_iq_j}(\vec{r}_{ij})=\left\{ \begin{array}{ll} 
\left[ q_iq_j=nn \right] \Rightarrow V_{\rm CON}(\vec{r}_{ij})+V_{\rm OGE}(\vec{r}_{ij})
+V_{\chi}(\vec{r}_{ij}) &  \\ 
\left[ q_iq_j=cn/cc \right]  \Rightarrow V_{\rm CON}(\vec{r}_{ij})+V_{\rm OGE}(\vec{r}_{ij}) &
\end{array} \right.\,,
\label{pot}
\end{equation}
where $n$ stands for the light quarks $u$ and $d$.
Notice that for the particular case of heavy quarks ($c$ or $b$) chiral symmetry is
explicitly broken and therefore boson exchanges associated to the dynamical breaking
of chiral symmetry do not contribute. The parameters of the model are the ones used 
for the study of the light one- and two-hadron systems~\cite{Val05,Vij05,Vac05,Val08,Gar99}, 
and for completeness they are quoted in Table~\ref{tab2}. 
\begin{table}[t]
\caption{Quark-model parameters.}
\label{tab2}
\begin{tabular}{lp{0.3cm}cp{0.5cm}|p{0.5cm}lp{0.3cm}c}
\hline
\hline
 $m_{u,d}$ (MeV)          && 313    &&& $g_{\rm ch}^2/(4\pi)$      && 0.54  \\ 
 $m_c$ (MeV)              && 1752   &&& $m_\sigma$ (fm$^{-1}$) && 3.42 \\ 
 $\hat r_0$ (MeV fm)      && 28.170 &&& $m_\pi$ (fm$^{-1}$)    && 0.70  \\ 
 $\mu_c$ (fm$^{-1}$)      && 0.70   &&& $\Lambda$ (fm$^{-1}$)  && 4.2  \\ 
 $b$ (fm)                 && 0.518  &&& $a_c$ (MeV)                && 230 \\ \hline\hline
\end{tabular}
\end{table}

\begin{figure}[t]
\vspace*{-1cm}
\includegraphics[width=.95\columnwidth]{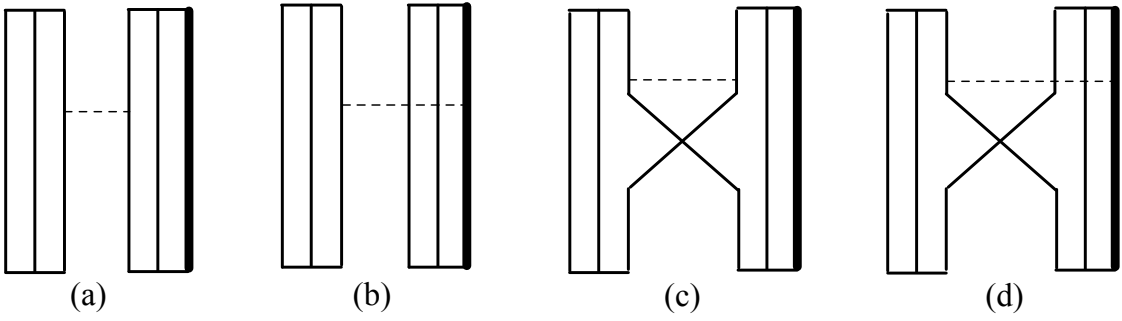}
\vspace*{-16.5cm}
\caption{Representative diagrams contributing to the charmed baryon$-$nucleon interaction.
The vertical solid lines represent a light quark, $u$ or $d$.
The vertical thick solid lines represent the charm quark. The dotted 
horizontal lines stand for the exchanged boson. (a) Interaction between two light quarks.
(b) Interaction between the heavy and a light quark. (c) Interaction between two light quarks
together with the exchange of identical light quarks. (d) Interaction between the heavy and a light quark
together with the exchange of identical light quarks.}
\label{fig1}
\end{figure}

In order to derive the $B_n B_m\to B_k B_l$ interaction from the
basic $qq$ interaction defined above, we use a Born$-$Oppenheimer
approximation where the quark coordinates are integrated out keeping $R$
fixed, the resulting interaction being a function of the two-baryon relative 
distance. A thorough discussion of the model can be found elsewhere~\cite{Val05,Vij05,Car15}. 
We show in Fig.~\ref{fig1} the different diagrams contributing to the charmed baryon$-$nucleon 
interaction. While diagrams (a) and (b) are considered in a hadronic description,
diagrams (c) and (d) correspond to short-range effects due to quark exchanges
that are not mapped in a hadronic description. Diagrams (c) and (d) contain 
one-gluon exchange contributions that are also missed in hadronic models. 
To illustrate the capability of the model let us just mention how the obtained 
$NN$ potentials perfectly describe the $S$ wave phase shifts~\cite{Gar99}.

In the limit where the two baryons $Y_c N$ overlap, the Pauli principle
may impose antisymmetry requirements not present in a hadronic description.
Such effects, if any, will be prominent for relative $S$ waves, $L=0$. The $S$ wave normalization 
kernel of the two-baryon wave function can be written in the overlapping region ($R \to 0$) 
as~\cite{Car15}
\begin{equation}
{\cal N}_{Y_c N}^{L=0SI} \stackrel[R\to 0]{}{\hbox to 20pt{\rightarrowfill}} 4\pi \left\lbrace {1 -\frac{R^2}{8} 
\left( \frac{5}{b^2} + \frac{1}{b_c^2}\right)}\right\rbrace  
\left\lbrace \left[ 1 - 3 C(S,I) \right] + ... \right\rbrace \, ,
\end{equation}
where $C(S,I)$ is a spin$-$isospin coefficient and $b$ and $b_c$
are the Gaussian parameters for the wave function of the light and charmed
quarks, respectively, assumed to be different for the sake of generality. 
The closer the value of $C(S,I)$ to 1/3 the larger 
the suppression of the normalization of the wave function 
at short distances, generating Pauli repulsion~\cite{Val97,Car15}.
Similarly to Pauli blocked channels, corresponding to $C(S,I)$=1/3, there might exist 
Pauli suppressed channels, those where $C(S,I)$ is close to $1/3$. 
This is the case for the channels $\Sigma_c N$ with $(I,J)=(1/2,0)$ and $(I,J)=(3/2,1)$ where $C(S,I)=8/27$
and $7/27$, respectively. The norm kernel gets rather small at 
short distances giving rise to Pauli repulsion. 
As we will discuss below, this repulsion will be reflected
in the phase shifts. Let us finally note that, although we will discuss 
the dependence of the results on different 
values of $b_c$, we take a reference value of $b_c=0{.}5$ fm.

\subsection{The coupled $\Lambda_c N - \Sigma_c N$ system}
\label{secIIb}

If we consider the system of two baryons $Y_c$ and $N$
in a relative $S$ state interacting through a potential $V$ that contains a
tensor force, then there is a coupling to the $Y_c N$ $D$ wave so that the
Lippmann$-$Schwinger equation of the system is of the form,
\begin{eqnarray}
t_{JI}^{\ell s\ell^{\prime \prime }s^{\prime \prime }}(p,p^{\prime \prime };E)
&=&V_{JI}^{\ell s\ell^{\prime \prime }s^{\prime \prime }}(p,p^{\prime \prime
})+\sum_{\ell^{\prime }s^{\prime }}\int_{0}^{\infty }{p^{\prime }}%
^{2}dp^{\prime }\,V_{JI}^{\ell s \ell^{\prime }s^{\prime }}
(p,p^{\prime })  \nonumber \\
&&\times {\frac{1}{E-{p^{\prime }}^{2}/2{\bf \mu }+i\epsilon }}%
t_{JI}^{\ell^{\prime }s^{\prime }\ell^{\prime \prime }s^{\prime \prime
}}(p^{\prime },p^{\prime \prime };E) \, ,  \label{eq1}
\end{eqnarray}
where $t$ is the two-body amplitude, $J$, $I$, and $E$ are the
total angular momentum, isospin and energy of the system, and $\ell s$, 
$\ell^{\prime }s^{\prime }$, $\ell^{\prime \prime }s^{\prime \prime }$ 
are the initial, intermediate, and final orbital angular momentum 
and spin. $p$ and $\mu $ are, respectively, the relative momentum and reduced mass of the
two-body system. More precisely, Eq.~(\ref{eq1}) is only valid
for the $\Sigma_c N$ system with isospin $3/2$. For this case, the
coupled channels of orbital angular momentum and spin
that contribute to a given state with total angular momentum $J$ 
are found in the first two rows of Table~\ref{tab1}.
\begin{table}[t]
\caption{$\Sigma_c N$ channels $(\ell_{\Sigma_c},s_{\Sigma_c})$ and 
$\Lambda_c N$ channels $(\ell_{\Lambda_c},s_{\Lambda_c})$ that contribute
to a given state with isospin $I$ and total angular momentum $J$.}
\begin{tabular}{cp{0.35cm}cp{1cm}cp{0.5cm}c}
\hline\hline
$I$ && $J$ && $(\ell_{\Sigma_c},s_{\Sigma_c})$ && $(\ell_{\Lambda_c},s_{\Lambda_c})$ \\ 
\hline 
3/2 && 0 && (0,0)       &&  \\ 
3/2 && 1 && (0,1),(2,1) && \\ 
1/2 && 0 && (0,0)       && (0,0)  \\ 
1/2 && 1 && (0,1),(2,1) && (0,1),(2,1) \\ \hline
\end{tabular}
\label{tab1}
\end{table}

In the case of isospin $1/2$, 
the $\Sigma_c N$ states are coupled to $\Lambda_c N$ states. Thus, 
if we denote the $\Sigma_c N$ system as channel $\Sigma_c$ and the 
$\Lambda_c N$ system as channel $\Lambda_c$, instead
of Eq.~(\ref{eq1}) the Lippmann$-$Schwinger equation for 
$\Lambda_c N -  \Sigma_c N$ scattering with isospin $1/2$ becomes,
\begin{eqnarray}
t_{\alpha\beta;JI}^{\ell_\alpha s_\alpha \ell_\beta s_\beta}(p_\alpha,p_\beta;E) & = & 
V_{\alpha\beta;JI}^{\ell_\alpha s_\alpha \ell_\beta s_\beta}(p_\alpha,p_\beta)+
\sum_{\gamma=\Lambda_c,\Sigma_c}\sum_{\ell_\gamma=0,2}
\int_0^\infty p_\gamma^2 dp_\gamma
V_{\alpha\gamma;JI}^{\ell_\alpha s_\alpha \ell_\gamma s_\gamma}(p_\alpha,p_\gamma)
 \nonumber \\ & & \times G_\gamma(E;p_\gamma) 
t_{\gamma\beta;JI}^{\ell_\gamma s_\gamma \ell_\beta s_\beta}
(p_\gamma,p_\beta;E);\,\,\,\,\, \alpha, \beta =\Lambda_c, \Sigma_c \,\,\, ,
\label{pup1}
\end{eqnarray}
where $t_{\Sigma_c \Sigma_c;JI}$ is the 
$\Sigma_c N \rightarrow \Sigma_c N$ scattering amplitude, 
$t_{\Lambda_c \Lambda_c ;JI}$ is the 
$\Lambda_c N \rightarrow \Lambda_c N$ scattering amplitude, 
and $t_{\Sigma_c \Lambda_c ;JI}$ is
the $\Sigma_c N \rightarrow \Lambda_c N$ scattering amplitude. 
The propagators $G_{\Lambda_c }(E;p_{\Lambda_c})$ and
$G_{\Sigma_c}(E;p_{\Sigma_c})$ 
in Eq.~(\ref{pup1}) are given by
\begin{eqnarray}
G_\Lambda(E;p_{\Lambda_c}) &= &\frac{2\mu_{N\Lambda_c}} 
{k_{\Lambda_c}^2-p_{\Lambda_c}^2+i\epsilon} \, ,
\label{pup2} \\
G_\Sigma(E;p_{\Sigma_c}) &= &\frac{2\mu_{N\Sigma_c}}{k_{\Sigma_c}^2-p_{\Sigma_c}^2+i\epsilon} \, ,
\label{pup3}
\end{eqnarray}
with
\begin{equation}
E=k_{\Lambda_c}^2/2\mu_{N\Lambda_c} \, ,
\label{pup3p}
\end{equation}
where the on-shell momenta $k_{\Sigma_c}$ and $k_{\Lambda_c}$ are related by
\begin{equation}
\sqrt{m_{N}^{2}+k_{\Lambda_c}^{2}}+\sqrt{
m_{\Lambda_c}^{2}+k_{\Lambda_c}^{2}}=  
\sqrt{m_{N}^{2}+k_{\Sigma_c}^{2}}+\sqrt{m_{\Sigma_c}^2+k_{\Sigma_c}^2} \, .
\label{forp10}
\end{equation}
We give in Table~\ref{tab1} the channels $(\ell_{\Lambda_c}, s_{\Lambda_c})$ 
and $(\ell_{\Sigma_c},s_{\Sigma_c})$,
corresponding to the $\Lambda_c N $ and $\Sigma_c N$ 
systems, which are coupled in a given state of total 
angular momentum $J$ for the case of isospin $1/2$.

\section{Results and discussion}
\label{secIII}
\subsection{$\Lambda_c N$ interaction}
\label{secIIIa}
\begin{figure}[t]
\vspace*{-1cm}
\includegraphics[width=.65\columnwidth]{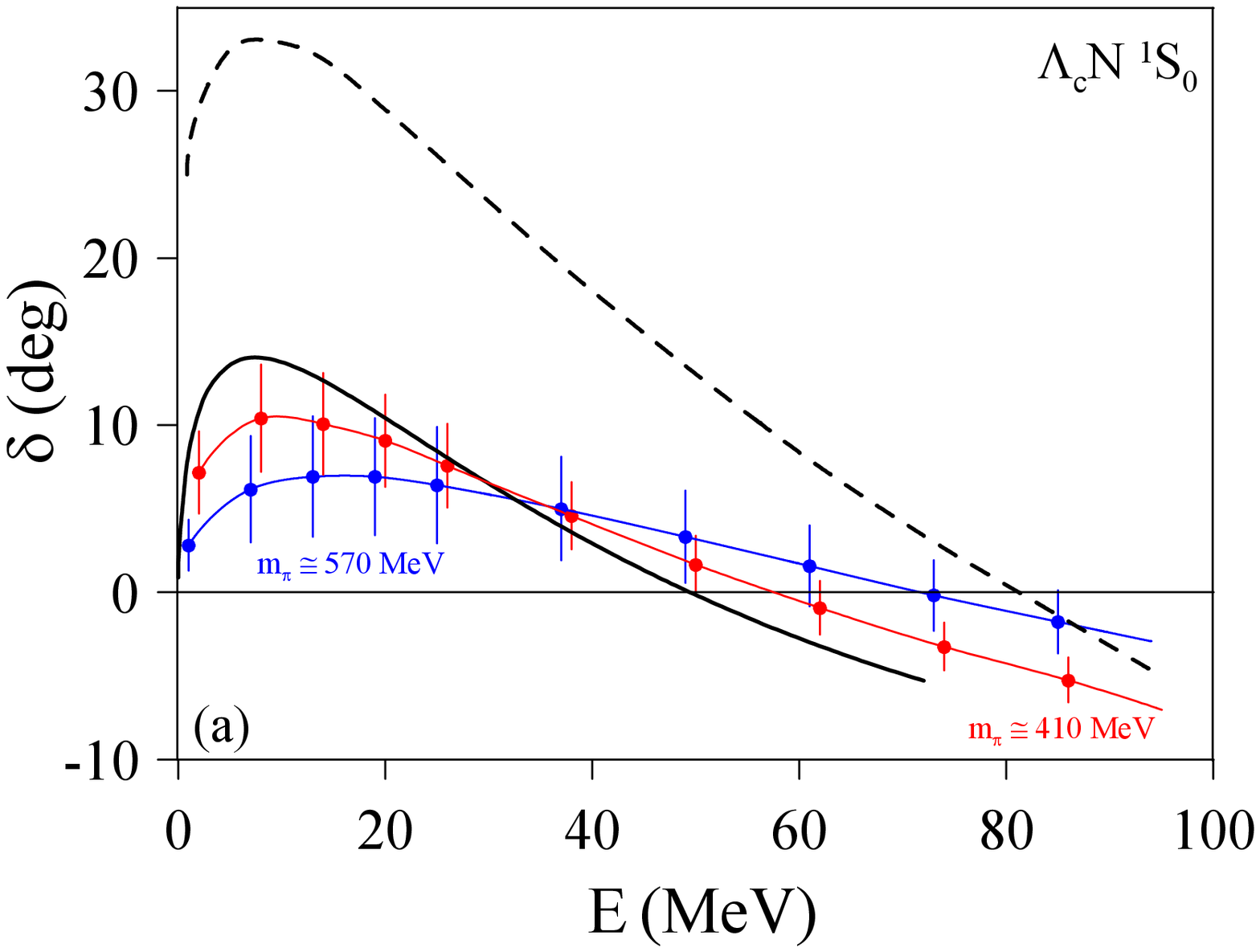}\vspace*{-8cm}
\includegraphics[width=.65\columnwidth]{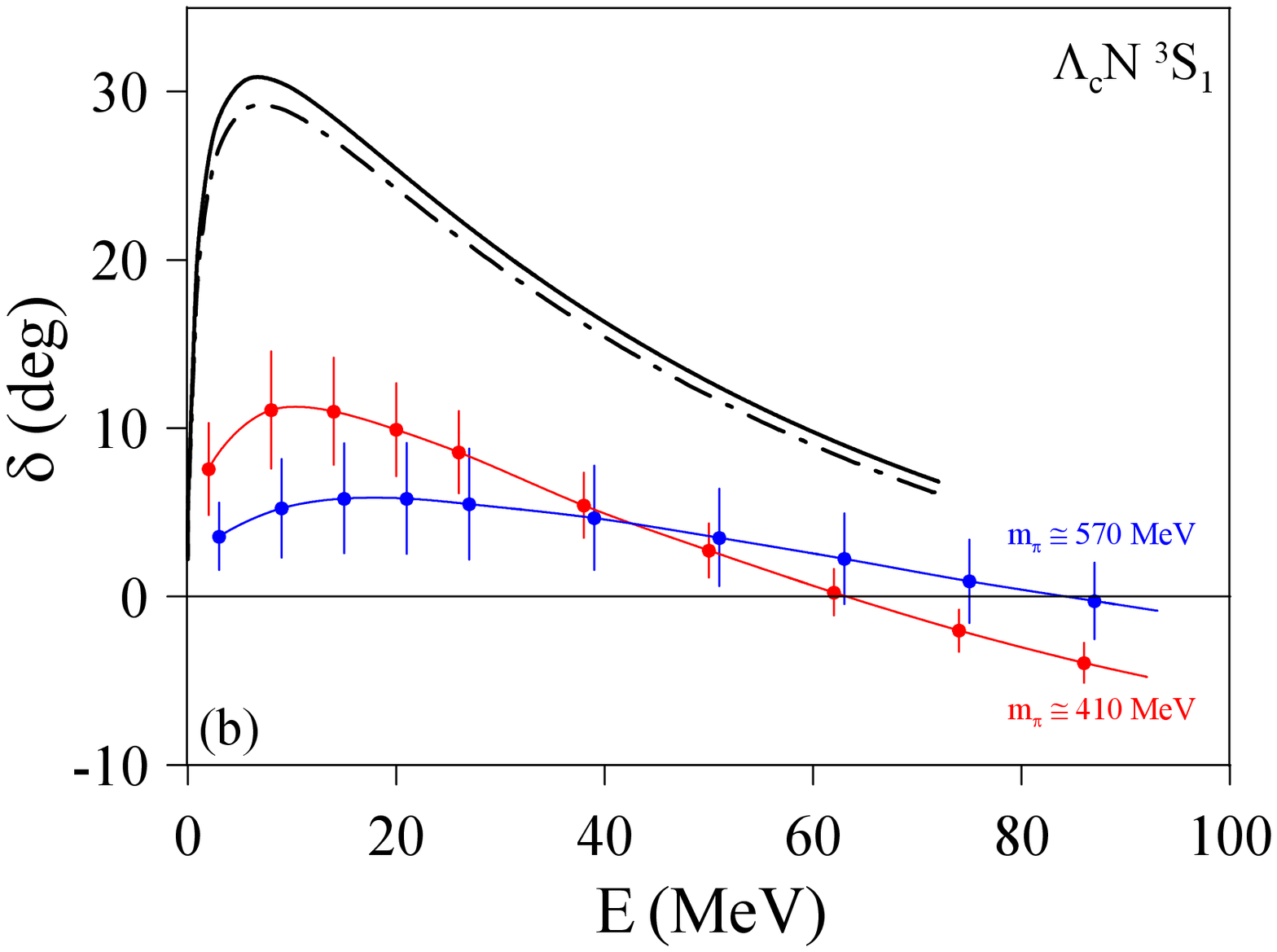}
\vspace*{-8.cm}
\caption{(a) Phase shifts for the $\Lambda_c N$ $^1S_0$ partial wave 
as a function of the c.m. kinetic energy.
The black solid line stands for the prediction of
the CQM. The black dashed line corresponds to the QDCSM results of Ref.~\cite{Hua13}.
The blue (red) filled circles represent the results of the 
HAL QCD Collaboration~\cite{Miy18} at $m_\pi =$ 570 (410) MeV. 
The vertical line at each point represents the statistical error of the lattice 
QCD simulations. The solid blue and red lines are just a guide to the eye.
(b) Phase shifts for the $\Lambda_c N$ $^3S_1$ partial wave 
as a function of the c.m. kinetic energy.
The black solid line stands for the prediction of the CQM. The black dashed-dotted 
line represents the CQM phase shifts without channel coupling.
The blue (red) filled circles represent the results of the 
HAL QCD Collaboration~\cite{Miy18} at $m_\pi =$ 570 (410) MeV.
The vertical line at each point represents the statistical error of the lattice 
QCD simulations. The solid blue and red lines are just a guide to the eye.}
\label{fig2}
\end{figure}
\begin{figure}[t]
\vspace*{-1cm}
\includegraphics[width=.65\columnwidth]{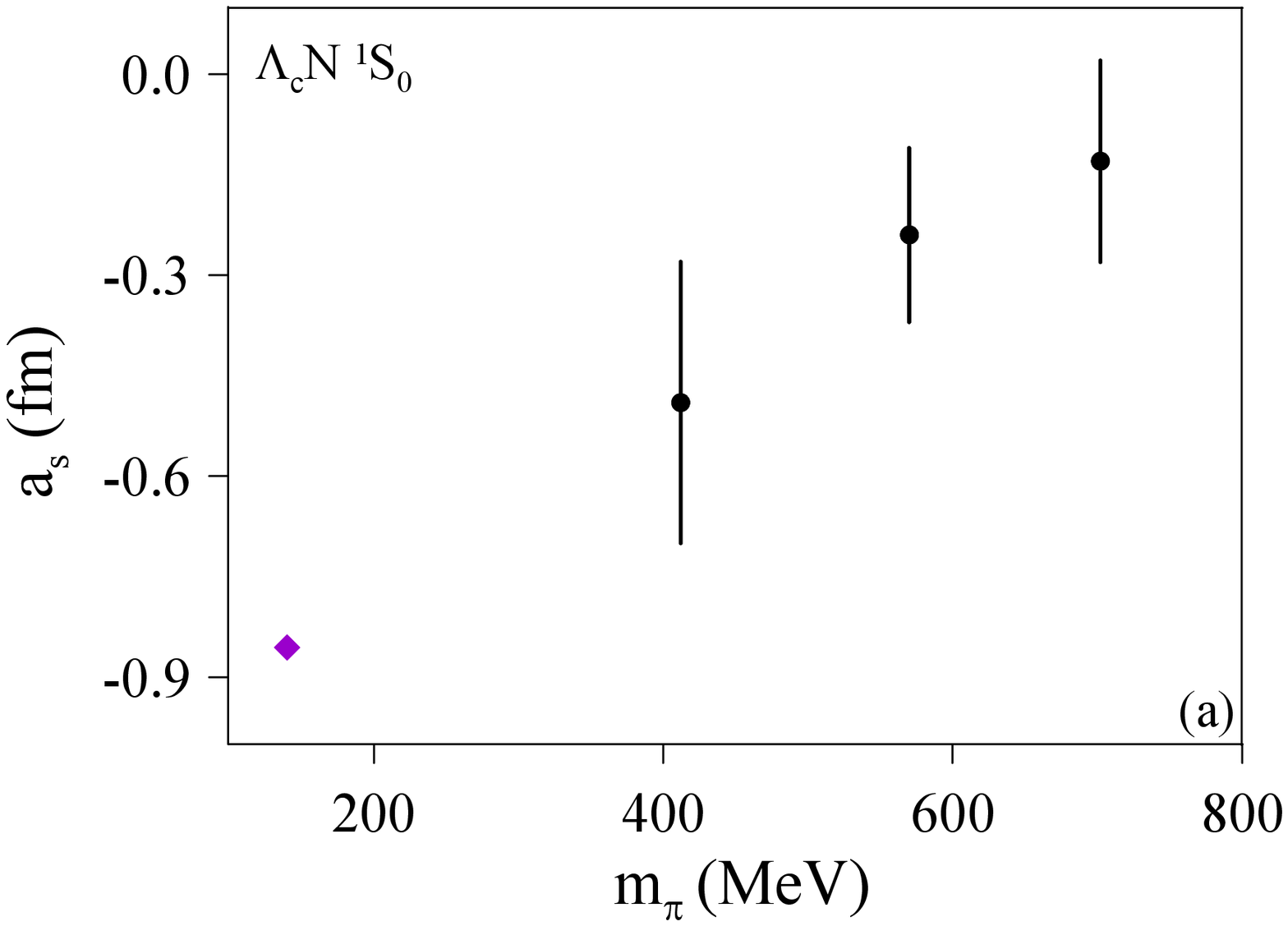}\vspace*{-8cm}
\includegraphics[width=.65\columnwidth]{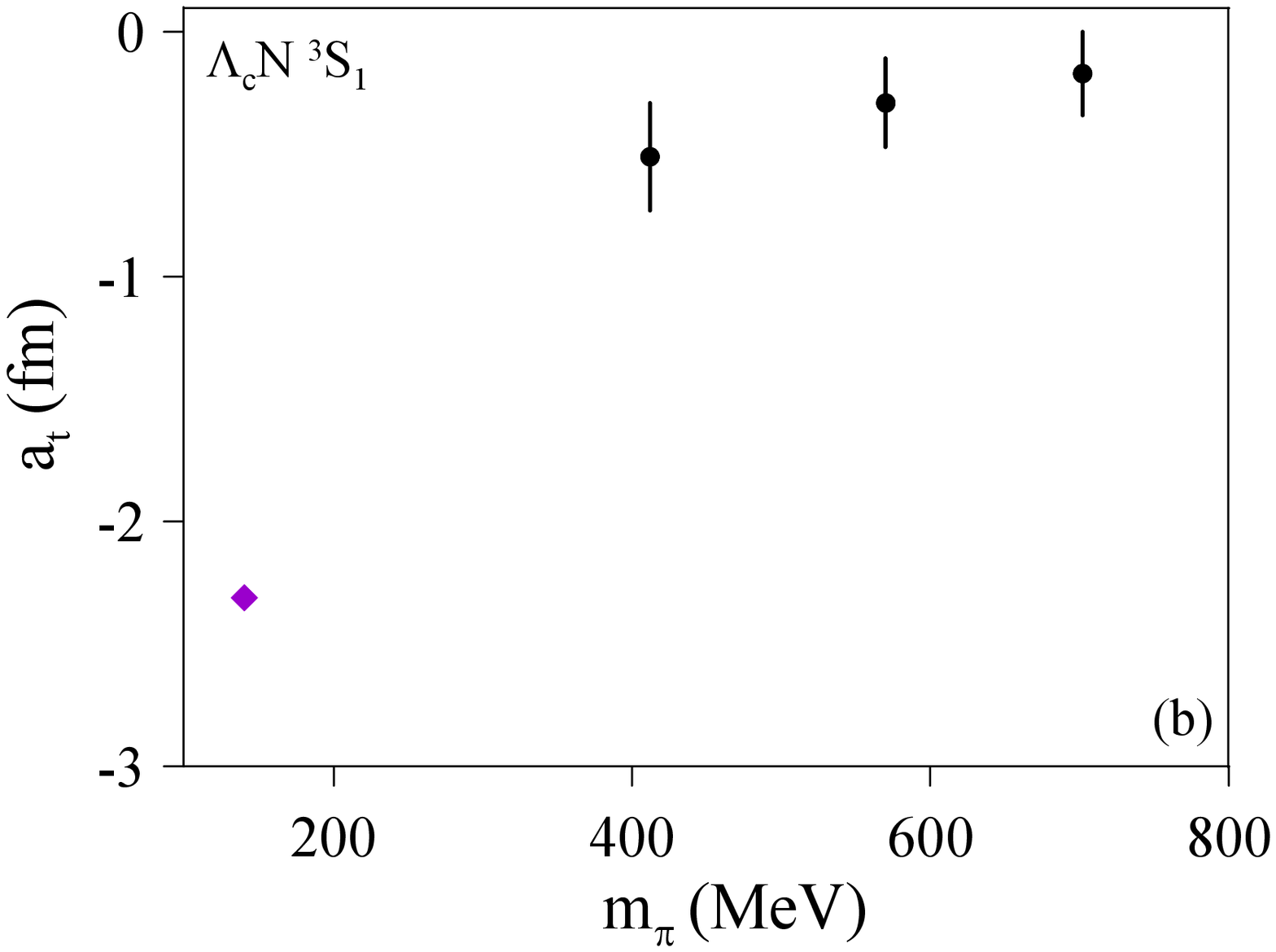}
\vspace*{-8.cm}
\caption{(a) Dependence of the HAL QCD $\Lambda_c N$ $^1S_0$ scattering length 
on the pion mass~\cite{Miy18}. The vertical bars include statistical and systematic errors.
The purple diamond represents the prediction of the CQM for the physical pion mass. 
(b) Same as (a) for the $\Lambda_c N$ $^3S_1$ partial wave.}
\label{fig3}
\end{figure}

We show in Fig.~\ref{fig2}(a) the phase shifts for the $\Lambda_c N$ $^1S_0$ partial wave
as a function of the center of mass (c.m.) kinetic energy. The
latest (2+1)-flavor lattice QCD simulations by the HAL QCD Collaboration~\cite{Miy18}\footnote{Preliminary
studies by the HAL QCD Collaboration of the $\Lambda_c N$ system~\cite{Miy15,Miy16} indicated an extremely 
weak interaction, while the latest results~\cite{Miy18} imply a somewhat stronger though still moderate 
attractive interaction.} for a pion mass of 570 (410) MeV are denoted by the blue (red) filled circles with their 
corresponding errors shown by the vertical lines. 
The black solid line stands for the results of the CQM described in Sec.~\ref{secIIa}. 
The black dashed line corresponds to the results of the QDCSM of Ref.~\cite{Hua13} for a color
screening parameter $\mu=0.1$. In Fig.~\ref{fig2}(b) we
present the phase shifts for the $\Lambda_c N$ $^3S_1$ partial wave {\textemdash} note that in
this case the results of the QDCSM model of Ref.~\cite{Hua13} are not available. 
As can be seen there is a tendency that the attraction obtained by the latest
lattice QCD simulations for both $\Lambda_c N$ $S$ waves becomes stronger as the pion mass
decreases, moving towards the predictions of the CQM and QDCSM models.

In Fig.~\ref{fig3} we show the dependence of the scattering lengths of the 
spin-singlet and spin-triplet $\Lambda_c N$ partial waves
reported by the HAL QCD Collaboration as a function of the pion mass. The purple diamonds at the
physical pion mass stand for the results of the CQM. The repulsive or attractive character of the interaction for the different
$Y_c N$ partial waves in the CQM is reflected in the scattering lengths and effective range parameters
summarized in Table~\ref{tab3}.

As can be seen in Figs.~\ref{fig2} and~\ref{fig3}, the phase shifts and scattering lengths 
of the $\Lambda_c N$ $^1S_0$ and $^3S_1$ partial waves derived by the HAL QCD Collaboration 
are qualitatively and quantitatively rather similar. Indeed, it was noted in Ref.~\cite{Miy18} that the corresponding $^1S_0$ and 
$^3S_1$ potentials are almost identical at 410 MeV pion mass and at 570 MeV. These potentials
show that the $\Lambda_c N$ interaction is attractive but not strong enough to form two-body bound 
states. The results of the CQM are slightly different: both partial
waves are attractive but without developing two-body bound states. However, the $^3S_1$ partial wave is more 
attractive than the $^1S_0$. This result is due to the short-range dynamics discussed in Sect.~\ref{secIIa},
consequence of gluon and quark exchanges. It has been outlined long ago in the literature 
for the $\Lambda N$ system~\cite{Fae87}. 

If no meson exchanges were considered, the $S$ wave phase shifts of the $\Lambda_c N$ system are 
very similar to the corresponding $NN$ scattering~\cite{Val95}. In both partial waves one
obtains typical hard-core phase shifts due to the short-range gluon and quark-exchange dynamics.
However, the hard-core radius in the spin-singlet state is larger than in the spin-triplet one~\cite{Fae87}
leading to a more attractive interaction in the spin-triplet partial wave due to a lower
short-range repulsion~\cite{Str88}. In fact, the hard cores caused by the color
magnetic part of the OGE potential have been calculated in Ref.~\cite{Fae87},
obtaining 0.35 fm for the spin-triplet state and 0.44 fm for the spin-singlet one. 
If the short-range dynamics is properly considered, this effect has to be 
transferred to the phase shifts, as concluded by the CQM. This difference stems from
the different expectation value in the spin-singlet and spin-triplet $\Lambda_c N$ partial waves 
of the color-magnetic operator appearing in Eq.~(\ref{OGE}),
$\vec\sigma_i \cdot \vec\sigma_j \vec\lambda^c_i \cdot \vec\lambda^c_j$. The matrix elements
of this operator are only different from zero when there are quark-exchange effects, as
depicted in diagrams (c) and (d) of Fig.~\ref{fig1}\footnote{If it were not so then
there would be net color exchange between two color singlet baryons, which is forbidden by QCD.},
giving rise to a genuine quark substructure effect not mapped at the hadronic level.
\begin{table}[t]
\caption{CQM results for the $^1S_0$ and $^3S_1$ scattering lengths ($\mathrm{a}_s$ and $\mathrm{a}_t$) and effective
range parameters ($r_s$ and $r_t$) in fm for the different $Y_c N$ systems.}
\begin{tabular}{cp{0.5cm}cp{1cm}cp{0.35cm}cp{1cm}cp{0.35cm}c} \hline\hline
$I$                    && System                   && $\mathrm{a}_s$         &&  $r_s$        &&  $\mathrm{a}_t$          &&  $r_t$  \\
\hline
\multirow{2}{*}{$1/2$} &&  $\Lambda_c N$           && $-$0.86                &&  5.64         &&  $-$2.31                 &&  2.97  \\
                       &&  $\Sigma_c N$            && $0{.}74 - i\, 0{.}18$  &&  $-$          &&  $-5{.}21 - i\, 1{.}96$  &&  $-$   \\   
$3/2$                  &&  $\Sigma_c N$            && $-$1.25                &&  8.70         &&  0.95                    &&  4.89  \\ \hline
\end{tabular}
\label{tab3}
\end{table}

Reference~\cite{Miy18} discusses the qualitative difference between the $\Lambda N$ and $\Lambda_c N$ 
interactions due to the absence of $K$-meson exchanges.
The origin of the small spin dependence of the $\Lambda_c N$ interaction
is attributed to the heavy $D$ meson mass and the large separation between the 
$\Lambda_c N$ and $\Sigma_c N$ masses. However, no discussion is found of the 
role of the short-range dynamics that may contribute 
to the different behavior of the spin-singlet and spin-triplet $\Lambda_c N$ phase shifts.
As will be discussed below, the short-range dynamics also generates a major impact
in the $\Sigma_c N$ charmed baryon$-$nucleon interaction. This is due to additional Pauli suppression,
as discussed in Sec.~\ref{secIIa}, in the $^1S_0 (I=1/2)$ and $^3S_1 (I=3/2)$ $\Sigma_c N$ partial waves, resulting
in a strong repulsion.

Recently, Ref.~\cite{Vid19} has presented a charmed baryon$-$nucleon
potential based on a $SU(4)$ extension of the meson-exchange hyperon-nucleon 
potential {\em \~A} of the J\"ulich group~\cite{Reu94}. Three different models of 
the interaction were considered, which differ only on the values of the couplings
of the scalar $\sigma$ meson with the charmed baryons. In particular, in a first 
model the couplings of the $\sigma$ meson with the charmed baryons are assumed to 
be equal to those of the $\Lambda$ and $\Sigma$ hyperons, and their values are taken from the original {\em YN}
potential {\em \~A} of the J\"ulich group. In the other two models these couplings are reduced
by 15\% and 20\%, respectively. The $\Lambda_c N$ phase shifts obtained with these models are
in qualitative agreement with the CQM results. They predict a higher overall attraction for the
$^3S_1$ than for the $^1S_0$ $\Lambda_c N$ partial wave, unlike the HAL QCD results, predicting similar phase 
shifts for both partial waves.

There are other studies of the $Y_c N$ interactions
based on one-boson exchange potentials at hadronic level~\cite{Liu12,Oka13}.
Although they do not report explicitly phase shifts or scattering lengths,
binding energies of the $Y_c N$ two-body systems as a function of the boson-exchange 
cutoff $\Lambda_\pi$ are calculated. As can be seen in Tables III and IX of Ref.~\cite{Liu12} the 
$J^P=0^+$ and $J^P=1^+$ states are bound for any value of $\Lambda_\pi$. The
binding energies of the $J^P=1^+$ state are always a little bit larger than those of the $J^P=0^+$ state. 
This is due to the similar contribution of the boson-exchange potentials in both
partial waves, the difference coming from the channel coupling that enhances the $D$ wave probability. 
Thus, while for $\Lambda_\pi=1.2$ GeV
the probability of the $^1S_0$ $\Lambda_c N$ channel in the $J^P=0^+$ state is 98.2\%, that of
the $^3S_1$ $\Lambda_c N$ channel in the $J^P=1^+$ state is 97.6\%, with a $D$ wave probability of 1.8\%. 
The small difference between the $^1S_0$ and $^3S_1$ probabilities in the $J^P=0^+$ and
$1^+$ states, remains
almost constant for any value of $\Lambda_\pi$. For example, for $\Lambda_\pi=1.6$ GeV they are
80.1\% and 79.6\%, respectively. However, the $D$ wave probability in the $J^P=1^+$ state
augments from 1.8 to 10.1\%. Table IV of Ref.~\cite{Liu12} reports 
binding solutions for the individual channels in the $J^P=0^+$ state.
As can be seen, the uncoupled $^1S_0$ $\Lambda_c N$ state is bound for any value of the
cutoff. Unfortunately, binding solutions for the uncoupled $\Lambda_c N$ channel in the $J^P=1^+$ 
state are not reported. A simplest guess-by-analogy estimation tells us that the results would be the
same in both $J^P$ states if channel coupling was not considered, as happens for the CQM if 
the short-range dynamics is neglected. 

In a later work~\cite{Mae16},
the hadron level one-boson exchange potential was supplemented by an overall short-range repulsion arising 
from color-magnetic effects evaluated in the heavy quark limit~\cite{Jaf77,Gig87,Lip87}. 
In general, the results are similar to their
previous study, both states $J^P=0^+$ and $J^P=1^+$ being bound or at the edge of 
binding and obtaining larger binding energies in the $1^+$ state for the same parametrization.
Hence, in both cases~\cite{Liu12,Oka13,Mae16} one expects phase shifts close to 180 degrees
at zero energy, being larger for the spin-triplet partial wave.

The phase shifts for the $^1S_0$ $\Lambda_c N$ interaction reported by the QDCSM model of
Ref.~\cite{Hua13}, dashed line in Fig.~\ref{fig2}(a), are more attractive than those of 
the CQM model, although they do not show a bound state. A major difference between the quark model
and hadron level approaches has to do with the strength of the channel coupling. The 
$\Lambda_c N - \Sigma_c N$ transition is rather weak both in the
quark-model description of Ref.~\cite{Hua13} and the hadronic or hybrid descriptions of
Refs.~\cite{Liu12,Oka13,Mae16,Vid19}. However, the tensor effects arising from the pseudoscalar
or vector meson exchanges become important at hadronic level\footnote{See for example
Table III of Ref.~\cite{Liu12} where binding energies on the order of hundred MeV are obtained for
the $J^P=0^+$ $\Lambda_c N$ state with a cutoff $\Lambda_\pi=1{.}7$ GeV.}, while they
are negligible in the QDCSM study of Ref.~\cite{Hua13}. We have calculated 
the $^3S_1$  $\Lambda_c N$ phase shifts with the CQM just by considering 
the diagonal interaction. The results are plotted by the 
dashed-dotted line in Fig.~\ref{fig2}(b), where the small contribution
of the channel coupling can be seen, in agreement with the QDCSM results of Ref.~\cite{Hua13}. It is worth to note that 
the $\Lambda_c - \Sigma_c$ conversion is less important than in the similar 
system in the strange sector, mainly due 
to the larger mass difference, namely 168~MeV as compared to 73 MeV in the strange sector.
Besides, it comes reduced as compared to the strange sector due to 
the absence of $K$-meson exchanges~\cite{Ban83}, generating a 
smaller $\Lambda_c N - \Sigma_c N$ transition potential.
The small contribution of the channel coupling obtained by the quark-model 
descriptions, CQM and QDCSM, to the charmed baryon$-$nucleon interaction
is in agreement with the observations of the HAL QCD
Collaboration, leading to the conclusion that the $\Lambda_c N$ tensor potential is negligibly weak~\cite{Miy18}
and that the coupling between $\Lambda_c N$ and $\Sigma_c N$ channels is also weak~\cite{Mia18}.
Similar conclusions were obtained in Ref.~\cite{Vid19}.
\begin{figure}[t]
\vspace*{-1cm}
\includegraphics[width=.65\columnwidth]{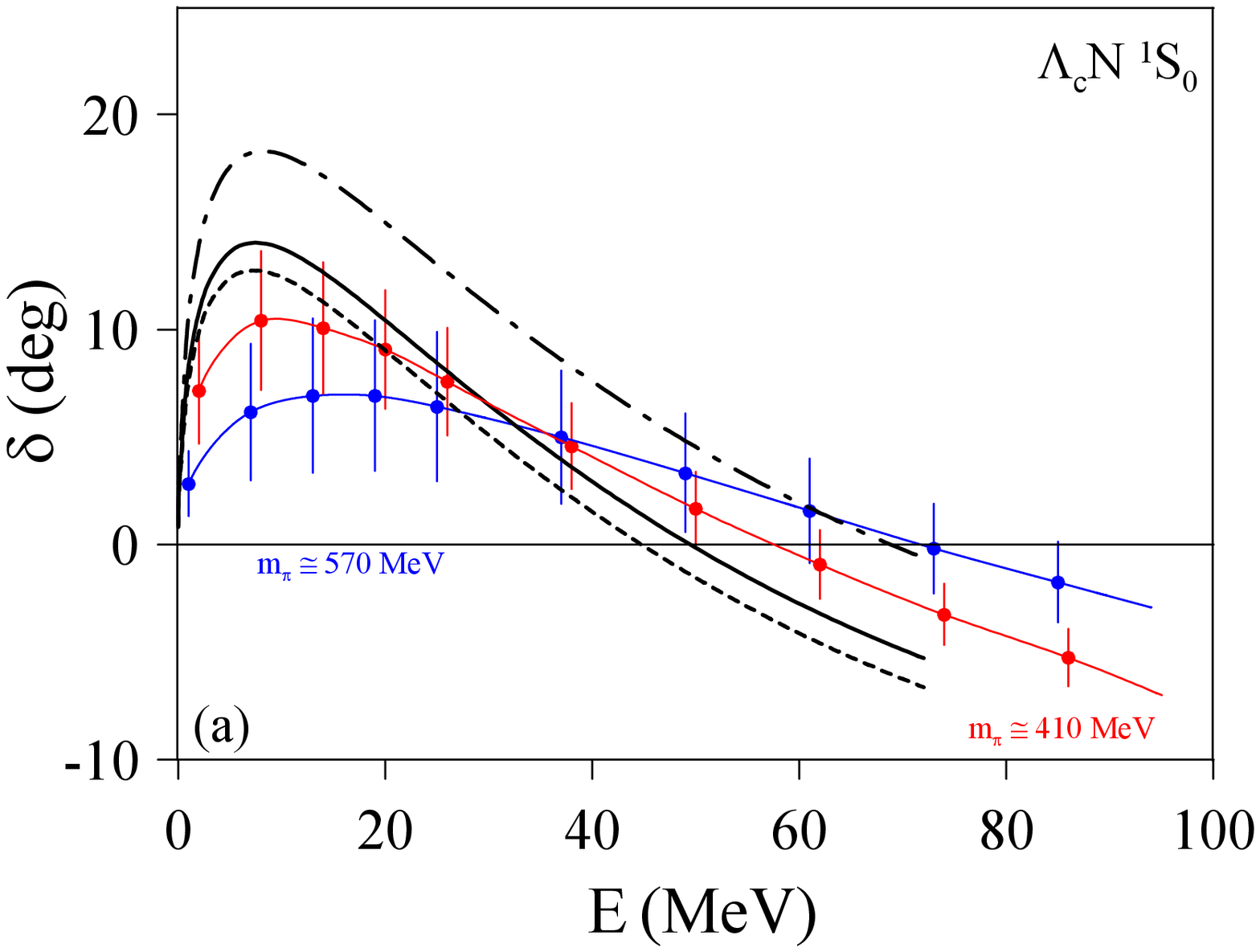}\vspace*{-8cm}
\includegraphics[width=.65\columnwidth]{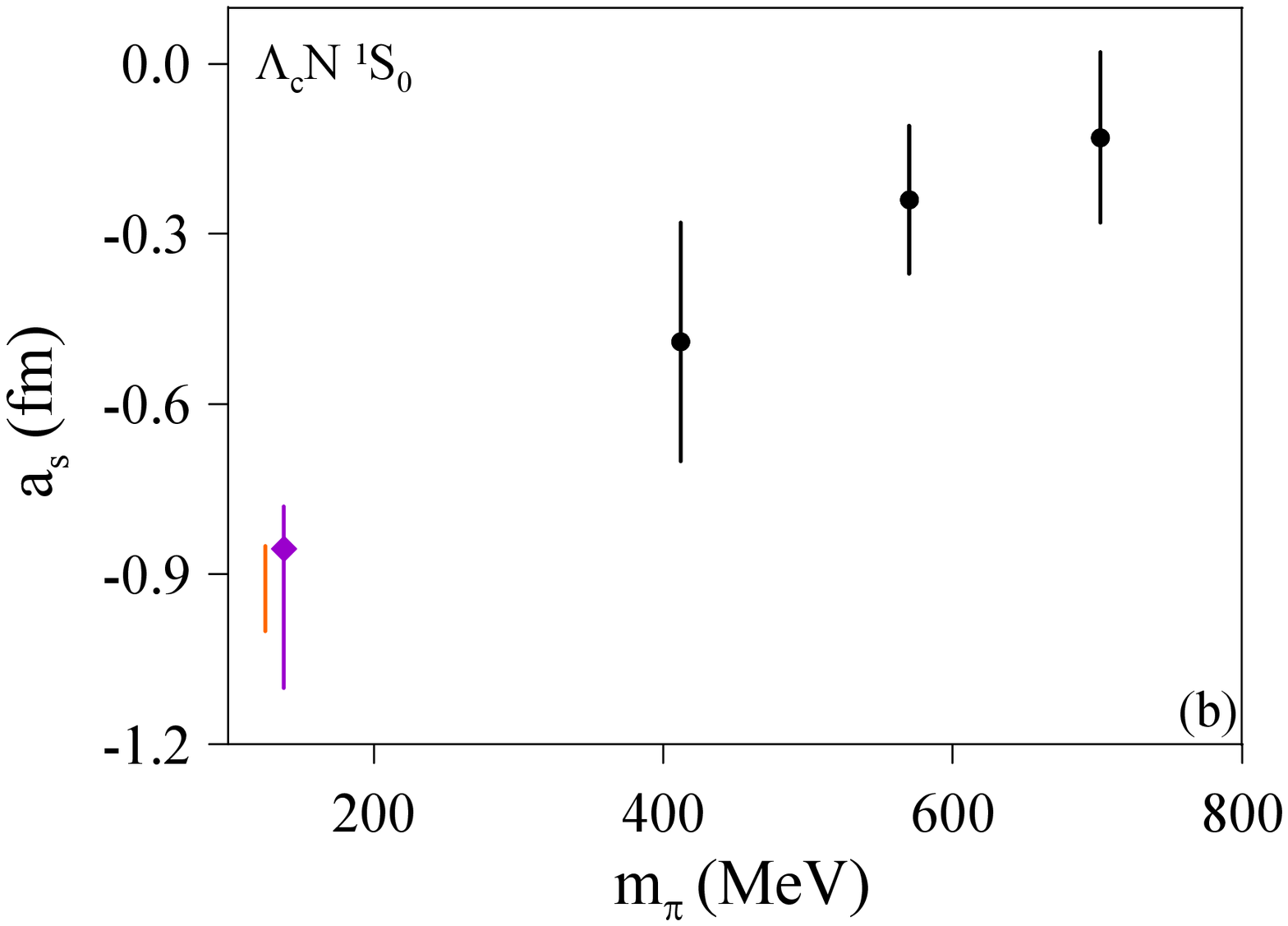}
\vspace*{-8.cm}
\caption{(a) Phase shifts for the $\Lambda_c N$ $^1S_0$ partial wave 
as a function of the c.m. kinetic energy.
The black solid line stands for the prediction of
the CQM for $b_c=0{.}5$ fm, the dashed-dotted for $b_c=0{.}2$ fm, and the dotted line for $b_c=0{.}8$ fm.
The blue (red) filled circles represent the results of the 
HAL QCD Collaboration~\cite{Miy18} at $m_\pi =$ 570 (410) MeV.
The vertical line at each point represents the statistical error of the lattice 
QCD simulation. The solid blue and red lines are just a guide to the eye.
(b) Dependence of the HAL QCD $\Lambda_c N$ $^1S_0$ scattering length 
on the pion mass~\cite{Miy18}. The vertical bars include statistical 
and systematic errors. The purple diamond represents the prediction 
for the physical pion mass of the 
CQM with the uncertainty associated to the range of values of $b_c$ 
chosen in (a). The orange vertical line stands for the range of values of the
EFT extrapolation of Ref.~\cite{Hai18} at the physical pion mass.}
\label{fig4}
\end{figure}

Reference~\cite{Hai18} has extrapolated the results of the HAL QCD Collaboration 
to the physical pion mass using EFT. The near-identity of the lattice QCD potentials extracted for the
$^1S_0$ and $^3S_1$ $\Lambda_c N$ partial waves~\cite{Miy18} persists in the extrapolation to the 
physical point. As the $^3S_1 - ^3{\!}D_1$ tensor coupling induced by the tensor forces
is taken into account in the EFT analysis of Ref.~\cite{Hai18}, it corroborates 
its smallness in the spin-triplet partial wave, as also derived from the CQM results of Fig.~\ref{fig2}(b). The EFT extrapolation
to the physical pion mass obtains a maximum  for the
$^1S_0$ $\Lambda_c N$ phase shift of around 17$-$21 degrees. This result is compatible with the predictions
of the CQM, as seen in Fig.~\ref{fig4}(a), where we have calculated the $^1S_0$ $\Lambda_c N$ phase
shifts for standard quark-model values of $b_c \in [0{.}2,0.{8}]$ fm. In Fig.~\ref{fig4}(b) we have
calculated the scattering length for the same interval of values of $b_c$ and we compare
with the result of the EFT extrapolation of Ref.~\cite{Hai18} at the physical pion mass, the orange vertical line,
getting also compatible results. The CQM predicts a slightly larger attraction for the
$^3S_1$ $\Lambda_c N$ partial wave. This result, which agrees with the
conclusions of Ref.~\cite{Vid19}, is not expected to coincide with the EFT 
extrapolation of the HAL QCD
$^1S_0$ and $^3S_1$ $\Lambda_c N$ phase shifts, $\mathrm{a}_t \in [-0{.}81,-0{.}98]$
fm, due to their identity at unphysical pion masses together with the already mentioned 
smallness of the tensor force in the spin-triplet partial wave. 
\begin{figure}[t]
\vspace*{-1cm}
\includegraphics[width=.65\columnwidth]{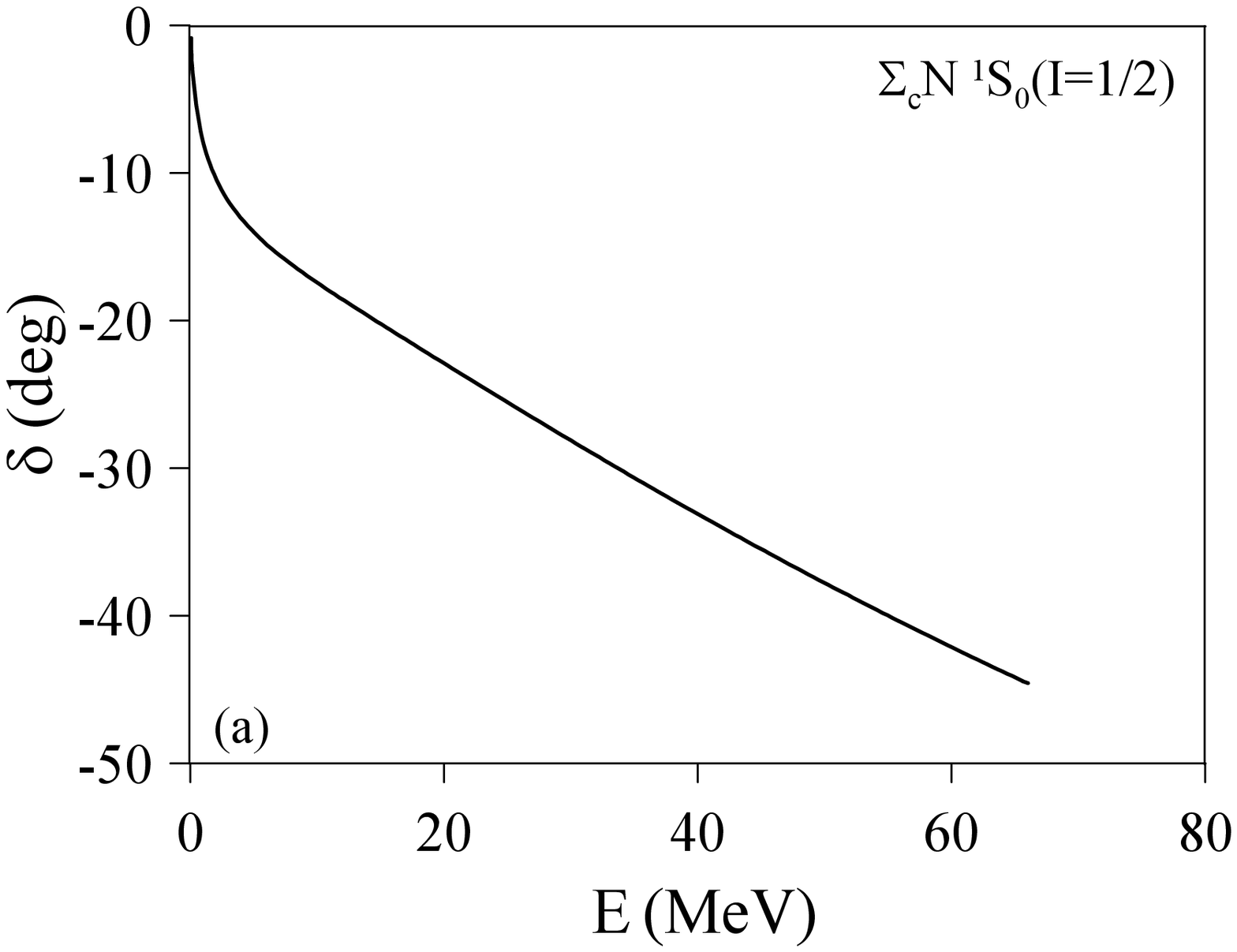}\vspace*{-8cm}
\includegraphics[width=.65\columnwidth]{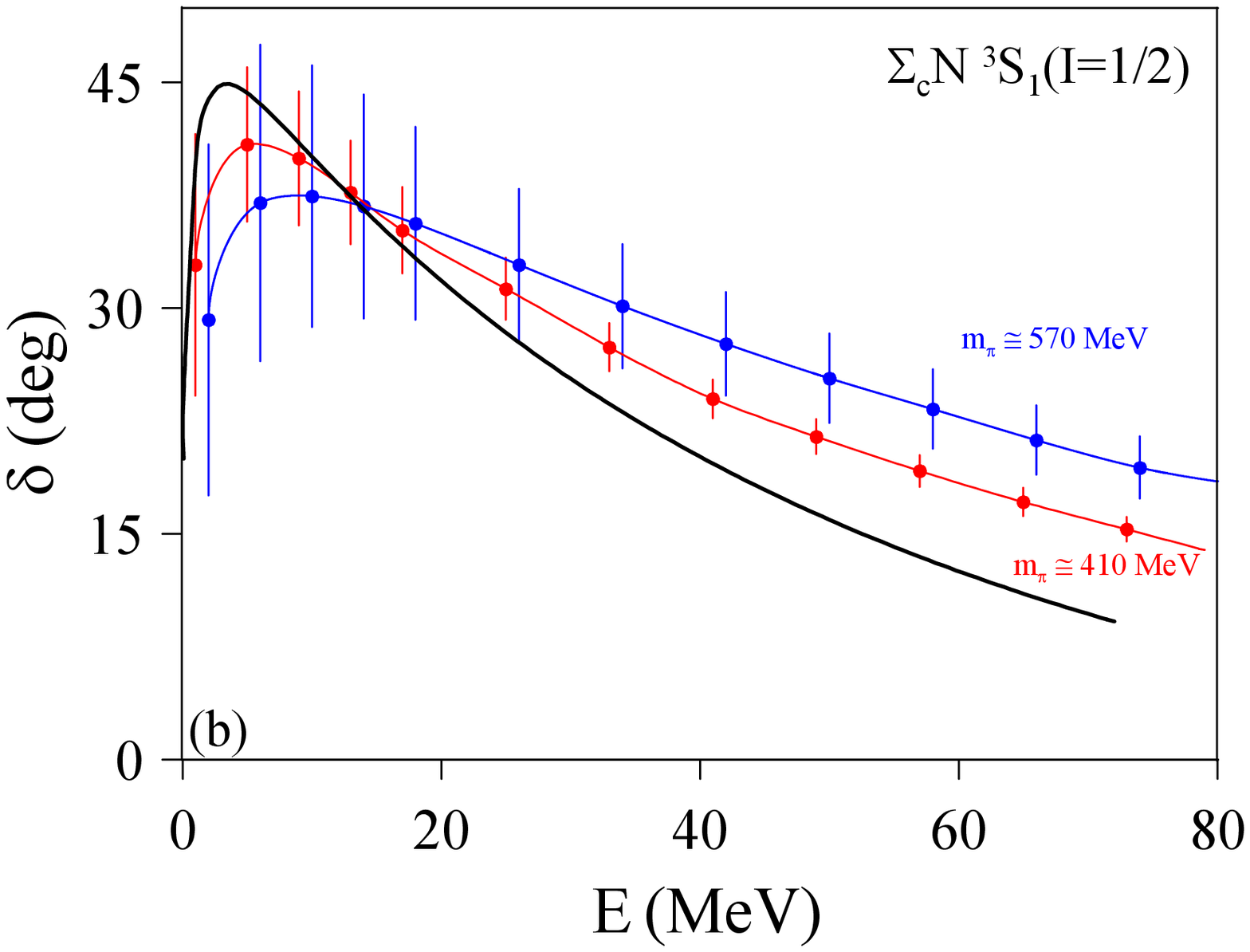}
\vspace*{-8.cm}
\caption{(a) Phase shifts for the $\Sigma_c N$ $^1S_0(I=1/2)$ partial wave,
as a function of the c.m. kinetic energy,
predicted by the CQM. (b) Phase shifts for the $\Sigma_c N$ $^3S_1(I=1/2)$ partial wave
as a function of the c.m. kinetic energy.
The black solid line stands for the prediction of the CQM.
The blue (red) filled circles represent the results of the 
HAL QCD Collaboration~\cite{Miy18} at $m_\pi =$ 570 (410) MeV.
The vertical line at each point represents the statistical error of the lattice 
QCD simulations. The solid blue and red lines are just a guide to the eye.}
\label{fig5}
\end{figure}
\begin{figure}[t]
\vspace*{-1cm}
\includegraphics[width=.65\columnwidth]{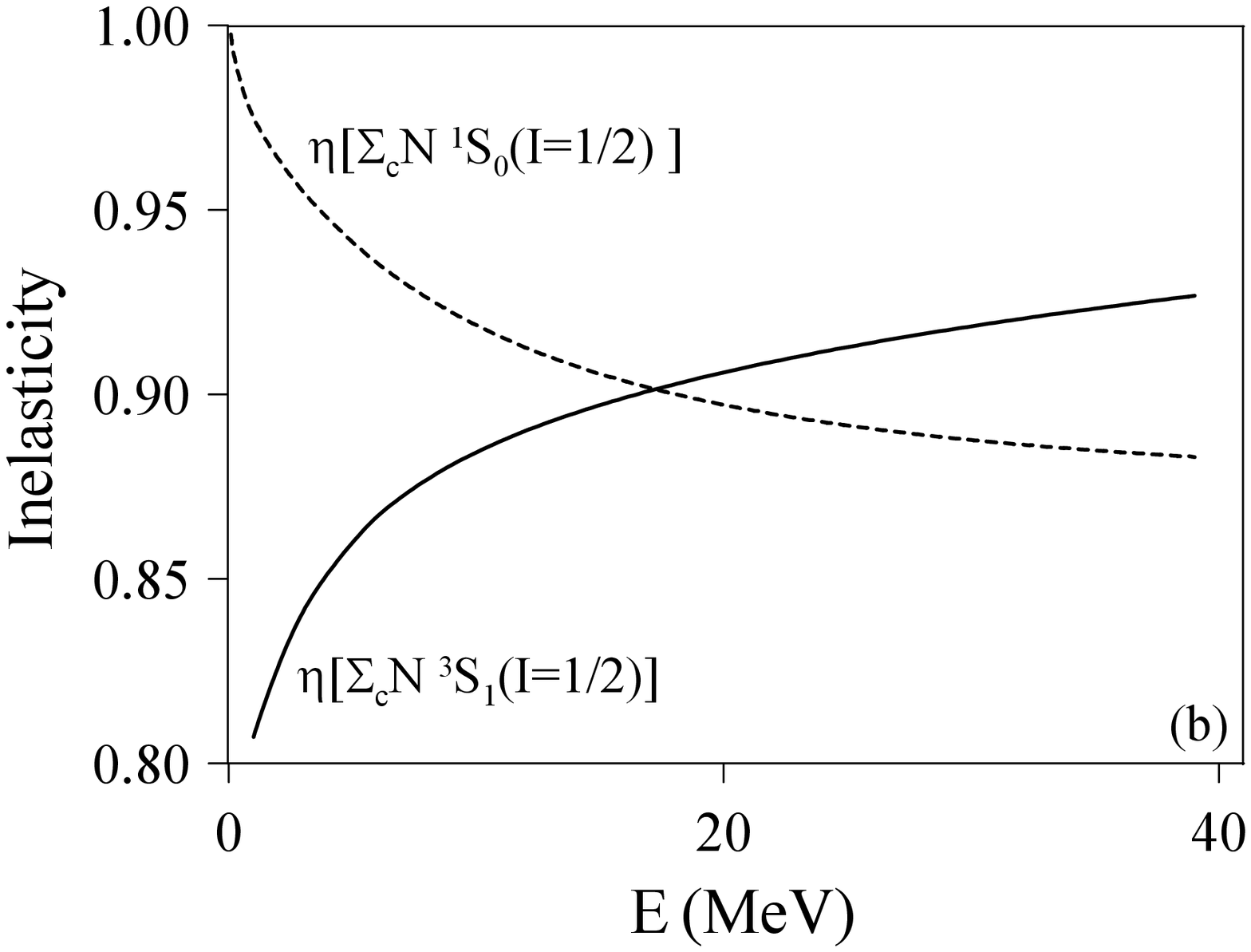}\vspace*{-8cm}
\includegraphics[width=.65\columnwidth]{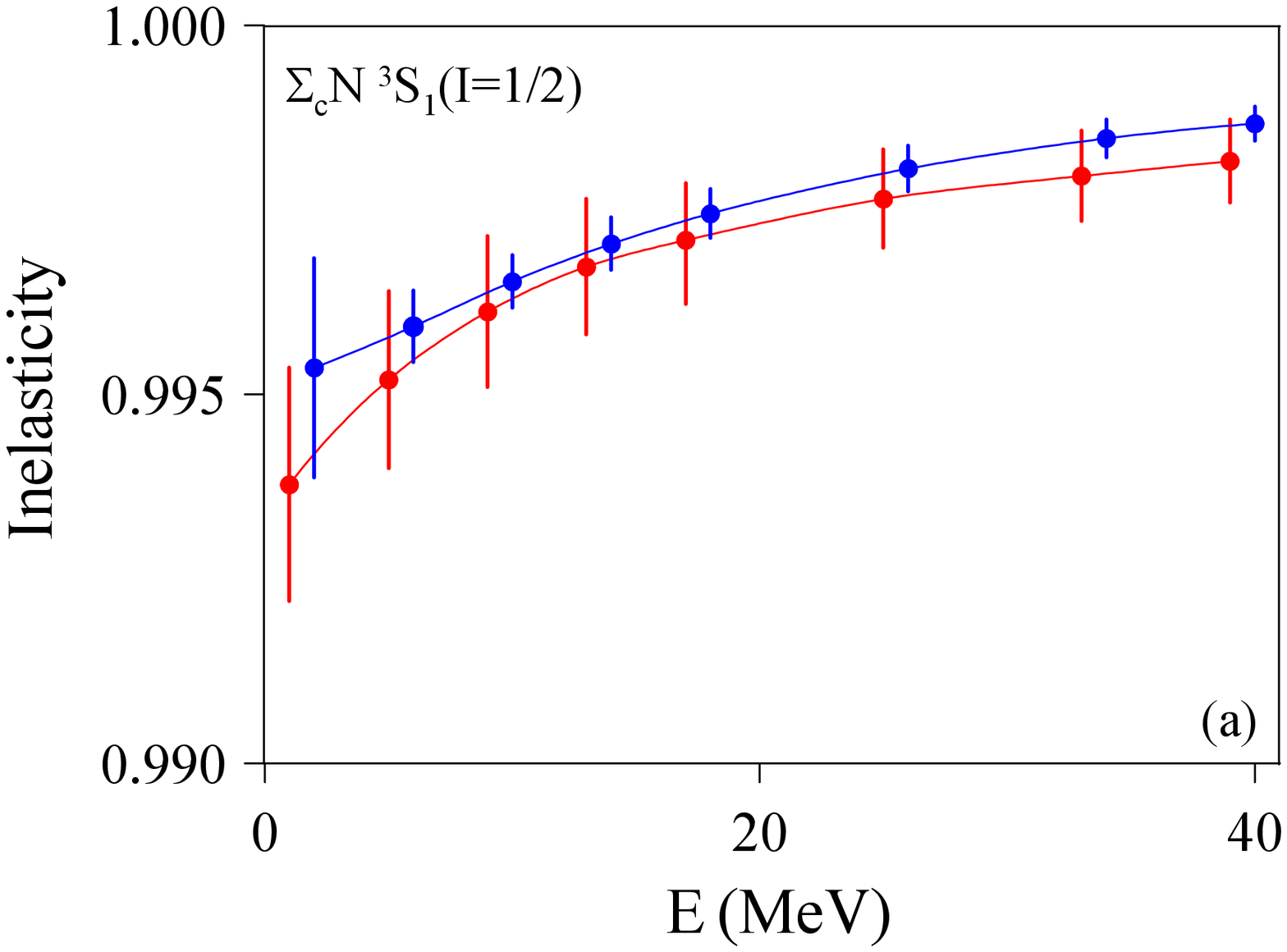}
\vspace*{-8.cm}
\caption{(a) Inelasticity for the $\Sigma_c N$ $^1S_0(I=1/2)$ 
and $\Sigma_c N$ $^3S_1(I=1/2)$ partial waves 
predicted by the CQM as a function of the c.m. kinetic energy.
(a) Inelasticity for the $\Sigma_c N$ $^3S_1(I=1/2)$ partial wave
of the HAL QCD Collaboration~\cite{Mia18} at $m_\pi =$ 570 MeV (blue filled circles) and
$m_\pi =$ 410 MeV (red filled circles) as a function of the c.m. kinetic energy.
The vertical line at each point represents the statistical error of the lattice 
QCD simulations. The solid blue and red lines are just a guide to the eye.}
\label{fig6}
\end{figure}
\begin{figure}[t]
\vspace*{-1cm}
\includegraphics[width=.65\columnwidth]{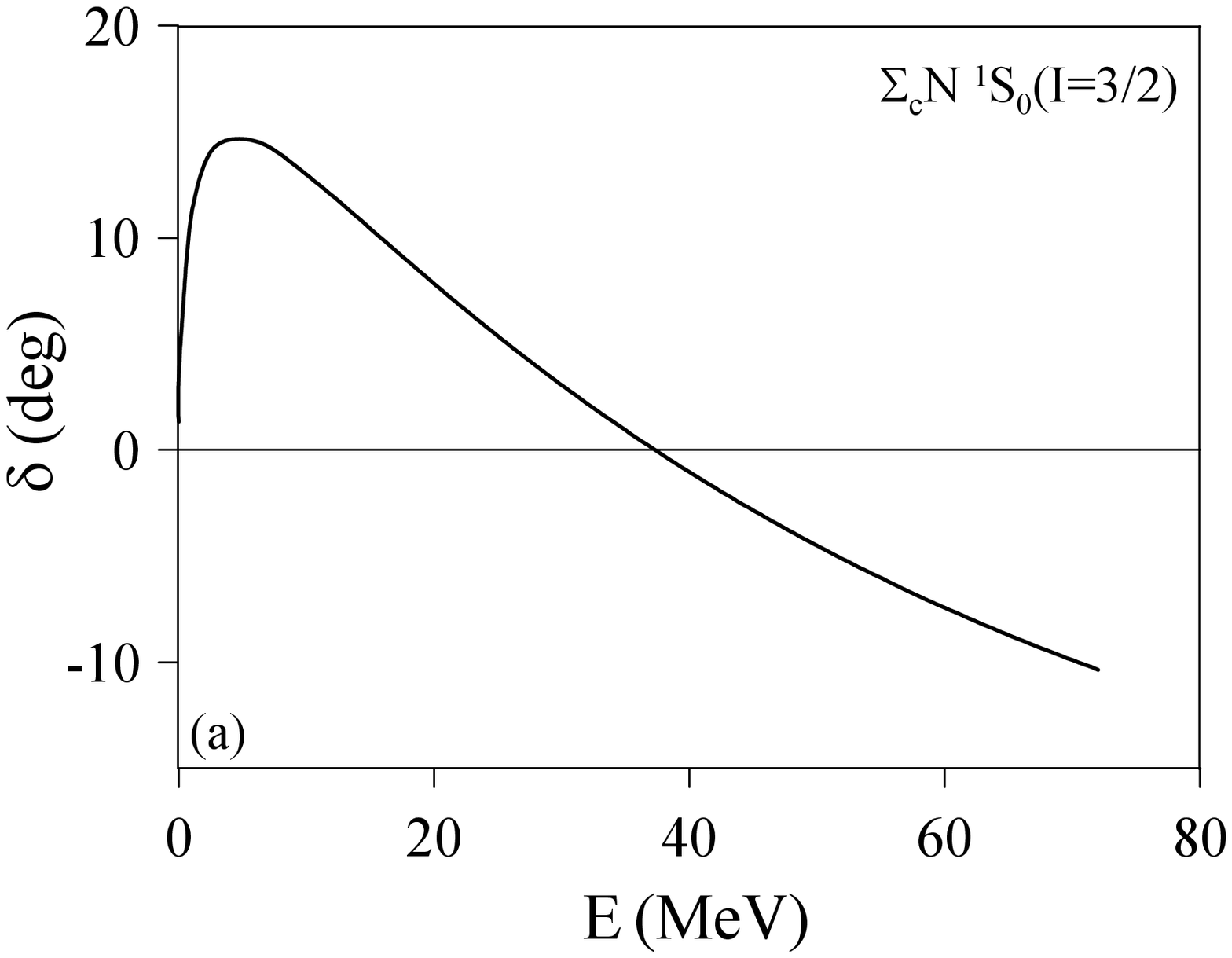}\vspace*{-8cm}
\includegraphics[width=.65\columnwidth]{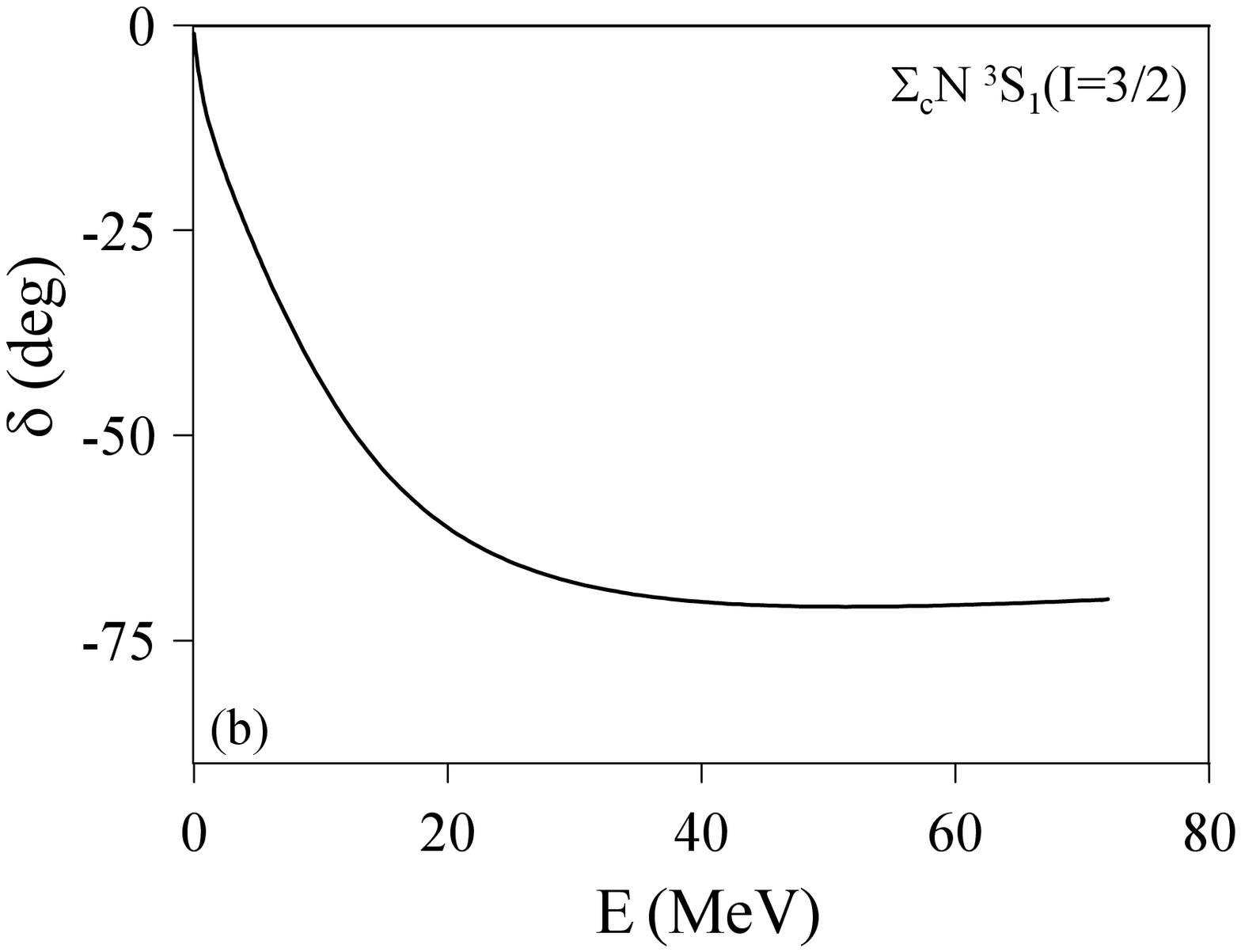}
\vspace*{-8.cm}
\caption{(a) Phase shifts for the $\Sigma_c N$ $^1S_0 (I=3/2)$ partial wave 
predicted by the CQM as a function of the c.m. kinetic energy. 
(b) Same as (a) for the $\Sigma_c N$ $^3S_1 (I=3/2)$ partial wave.}
\label{fig7}
\end{figure}

\subsection{$\Sigma_c N$ interaction}
\label{secIIIb}

In Fig.~\ref{fig5} we show the $I=1/2$ $\Sigma_c N$ phase shifts. Figure~\ref{fig5}(a)
presents the prediction of the CQM model for the $^1S_0$ partial wave. 
There are no data available to compare with.
The strong repulsion observed in the $\Sigma_c N$ $^1S_0 (I=1/2)$ interaction
is a consequence of Pauli suppression effects arising in spin$-$isospin saturated channels~\cite{Val97},
as discussed in Sect.~\ref{secIIa}. Results of other theoretical approaches for this partial
wave would help to disentangle the role of the short-range dynamics 
in the charmed baryon$-$nucleon interaction.
Figure~\ref{fig5}(b) shows the phase shifts for the $^3S_1$ partial wave. 
The black solid line stands for the results of the CQM.
The latest (2+1)-flavor lattice QCD simulations by the HAL QCD Collaboration~\cite{Miy18}
for a pion mass of 570 (410) MeV are shown by the blue (red) filled circles with their corresponding 
errors. As in the $\Lambda_c N$ interaction, the tendency can be seen 
that the attraction becomes stronger as the pion mass
decreases, the phase shifts moving towards the results of the CQM. 
One observes that lattice QCD simulations predict the attraction in the $\Sigma_c N$ $^3S_1 (I=1/2)$ channel 
to be stronger than in the equivalent $\Lambda_c N$ channel. This conclusion also holds for the CQM 
results. The attractive or repulsive character of the $\Sigma_c N$ interaction in the CQM is reflected in the 
scattering lengths given in Table~\ref{tab3}. Note that the scattering lengths 
of the $I=1/2$ $\Sigma_c N$ system are complex because the lower $\Lambda_c N$ channel
is always open.

Figure~\ref{fig6}(a) shows the inelasticity for the $\Sigma_c N$ $^3S_1 (I=1/2)$ partial wave
derived by the HAL QCD Collaboration~\cite{Mia18}
for a pion mass of 570 (410) MeV by blue (red) filled circles with their corresponding 
errors. Figure~\ref{fig6}(b) shows the inelasticity obtained with the CQM 
for the $\Sigma_c N$ $^3S_1$ 
and $^1S_0$ $I=1/2$ partial waves. Although the coupling between the $\Lambda_c N$ and $\Sigma_c N$
channels in the $^3S_1$ partial wave is small, see Fig.~\ref{fig2}(b), the inelasticity
predicted by the CQM is larger than the HAL QCD simulation.
\begin{figure}[t]
\vspace*{-1cm}
\includegraphics[width=.65\columnwidth]{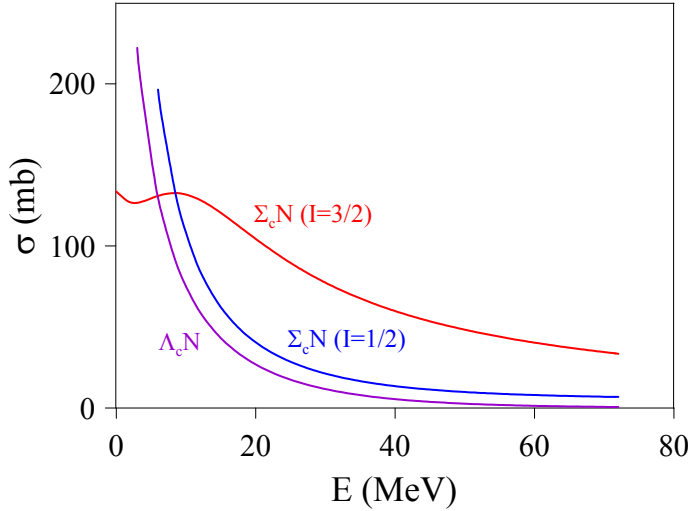}
\vspace*{-8.cm}
\caption{Total cross section for the $\Lambda_c N$ (purple line),
$\Sigma_c N (I=1/2)$ (blue line) and $\Sigma_c N (I=3/2)$ (red line)
scattering. $E=0$ stands for the $\Lambda_c N$ or $\Sigma_c N$ thresholds.}
\label{fig8}
\end{figure}

In Fig.~\ref{fig7} we show the $I=3/2$ $\Sigma_c N$ phase shifts.
The $^1S_0$ $\Sigma_c N$ channel presents an attraction comparable to the $^1S_0$ $\Lambda_c N$ 
system. The scattering length is still far
from the standard values of the $\Lambda N$ system, in the order of $-2{.}9$ to $-2{.}6$ fm,
which may allow for the existence of three-body bound states as we will discuss below.
The $^3S_1 (I=3/2)$ $\Sigma_c N$ channel presents a strong repulsion, a consequence 
again of quark-Pauli effects arising in spin$-$isospin saturated channels.
As mentioned above for the $\Sigma_c N$ $^1S_0 (I=1/2)$ state, it would be 
convenient to have results of other theoretical approaches for the phase shifts of
the $\Sigma_c N$ $^3S_1 (I=3/2)$ partial wave, to scrutinize the short-range
dynamics.

Finally, in Fig.~\ref{fig8} we present the CQM results for the total cross section 
for the $\Lambda_c N$, $\Sigma_c N (I=1/2)$, and $\Sigma_c N (I=3/2)$
scattering.

\subsection{$\Lambda_c$ hypernuclei.}
\label{secIIIc}

One of the most interesting applications of the charmed baryon$-$nucleon interaction
is the study of the possible existence of charmed hypernuclei. The binding energy
of $\Lambda_c$ hypernuclei has been analyzed in Ref.~\cite{Miy18}
using the HAL QCD $\Lambda_c N$ interaction for $m_\pi= 410$ MeV, where it has been noted that
for nuclei with $A=12-58$ the Coulomb repulsion is not much stronger than
the strong binding energy, which leads to the possible existence of $\Lambda_c$ 
hypernuclei in light or medium$-$heavy nuclei. On the contrary,
Refs.~\cite{Gar15,Mae16} concluded to the existence of light $\Lambda_c$ hypernuclei.
Moreover, Ref.~\cite{Vid19} concluded to the existence of $\Lambda_c$ hypernuclei
for all nuclei studied, from $^5$He to $^{209}$Pb.

Regarding the possibility of a $J=1/2$ charmed hypertriton with the HAL QCD $Y_c N$
interactions there is a delicate balance. On the one hand, it would be favored, bearing 
in mind the tendency that the $\Lambda_c N$ attraction becomes stronger as the 
pion mass decreases. On the other hand, since the average $\Lambda_c N$ ($\Lambda N$) 
potential that it is relevant for the charmed hypertriton (hypertriton) is dominated 
by the spin-singlet channel~\cite{Gib94},
the considerably smaller $^1S_0$ $\Lambda_c N$ scattering length compared to the
$\Lambda N$ system goes against its existence. 
The balance could be tilted if the spin dependence of the $\Lambda_c N$ interaction
induced by the short-range dynamics would slightly enhance the 
attraction in the spin-triplet partial wave as compared with the spin-singlet one. Then, 
the existence of $J=3/2$ $\Lambda_c$ hypernuclei might be considered seriously.
The isoscalar $J=3/2$ state is dominated by the more attractive 
spin-triplet interaction~\cite{Miy95},
which together with the reduction of the kinetic energy associated with
the $\Lambda_c$ induced by its larger mass as compared to
the $\Lambda$, could lead to a slightly bound $J=3/2$ $\Lambda_c$ hypernucleus~\cite{Gar15}.
In this regard, it is important to keep in mind
that the isoscalar $J=3/2$ $\Lambda NN$ state 
is close to threshold; see Table V and Fig. 2 of Ref.~\cite{Gar07}.

The recent few-body calculation of Ref.~\cite{Mae16}, employing the strongly
attractive one-boson exchange interactions discussed above leading already to 
$\Lambda_c N$ bound states, leads to several $\Lambda_c NN$ bound states with 
binding energies of the order of 20 MeV. As has been discussed above, one
made use of a slightly more attractive interaction for the $\Lambda_c N$ spin-triplet
partial wave than for the spin-singlet partial wave.
This generates an isoscalar $J=3/2$ $\Lambda_c NN$ ground state 
instead of $J=1/2$, see Fig. 11 of Ref.~\cite{Mae16}. 

The order of the isoscalar $\Lambda_c NN$ $J=1/2$ and $J=3/2$ channels
is also reversed with respect to the strange
sector in the CQM model, the $J=3/2$ being the most attractive one. This difference
can easily be associated with the importance of the $\Lambda - \Sigma$
conversion in the strange sector~\cite{Miy95}. 
When the $\Lambda N - \Sigma N$ potential
is disconnected, the $J=3/2$ channel is almost not modified, while the $J=1/2$
loses great part of its attraction. Thus, the ordering between the $J=1/2$ and $J=3/2$ 
channels is reversed in such a way that the hypertriton would
not be bound (see Fig. 6(a) of Ref.~\cite{Gar07}). 
As we have already discussed, the $\Lambda_c - \Sigma_c$ conversion is less
important than in the strange sector, giving rise to a 
softer $\Lambda_c N - \Sigma_c N$ transition potential.
Thus, the calculation of Ref.~\cite{Gar15} making use of the
CQM phase shifts presented in Figs.~\ref{fig2} and~\ref{fig5}, i.e.,
without two-body bound states, 
obtained an isoscalar $J=3/2$ charmed hypernucleus with a 
binding energy of 0.27 MeV. After correcting exactly by the Coulomb
potential the final binding energy obtained was 0.14 MeV. Different from the
hadron level calculation of Ref.~\cite{Mae16}, in the CQM model the $J=1/2$
$\Lambda_c NN$ is unbound. Let us finally note that the hard-core radius of 
the $\Lambda_c N$ interaction, relevant for the study of charmed hypernuclei~\cite{Ban83}, 
in the CQM is fixed by the short-range dynamics~\cite{Fae87}.

There are not few-body calculations with the
QDCSM $Y_c N$ interactions of Ref.~\cite{Hua13}. However, a simple reasoning 
hints towards the possible existence of a $J=1/2$ charmed hypertriton in this 
model. As one can see in Fig.~\ref{fig2}(a), the $\Lambda_c N$ $^1S_0$ phase shifts
predicted by the QDCSM are similar to those of the $\Lambda_c N$ $^3S_1$ 
partial wave obtained with the CQM, see Fig.~\ref{fig2}(b). As 
the channel coupling is negligible in both cases, with the QDCSM one would obtain 
a scattering length for the $\Lambda_c N$ $^1S_0$ state of about 
$-$2.31 fm, see Table~\ref{tab2}. This scattering length is within the order of 
that of the $^1S_0$ $\Lambda N$ system, between $-$2.9 and $-$2.6 fm, which is
a key ingredient for the existence of the hypertriton. 
The possible existence of a $J=1/2$
charmed hypertriton in the QDCSM would be reinforced
by the reduced kinetic energy contribution of the $\Lambda_c$ baryon. 
It might be at an disadvantage by the lack of the $\Lambda N - \Sigma N$
coupling that, as seen in Fig. 6(a) of Ref.~\cite{Gar07}, is of basic 
importance to get the hypertriton in quark-model based descriptions.

Reference~\cite{Vid19} has also studied the possible existence of
bound states of the $\Lambda_c$ in different nuclei. One makes use of the
$\Lambda_c$ self-energy as an effective $\Lambda_c$-nucleus mean-field potential 
in a Schr\"odinger equation to get the bound state energies. $\Lambda_c$ hypernuclei
from $^5_{\Lambda_c}$He to $^{209}_{\Lambda_c}$Pb are studied. Even the less
attractive model for the $Y_c N$ interaction of those discussed in Sect.~\ref{secIIIa},
where the couplings of the $\sigma$ meson with the charmed baryons are reduced
20\% as compared to the original {\em YN} potential {\em \~A} of the J\"ulich group, 
is able to bind the $\Lambda_c$ in all the nuclei considered. This is in contrast 
with the HAL QCD Collaboration results~\cite{Miy18}, which suggest that only
light- or medium-mass $\Lambda_c$ nuclei could really exist. The conclusions of this
work come to reinforce the results obtained with the CQM in Ref.~\cite{Gar15}. On the 
one hand they arrive at the same conclusion as regards the negligible contribution of the
$\Lambda_c N -\Sigma_c N$ coupling, and on the other hand they support 
the possible existence of light charmed hypernuclei.

\section{Outlook}
\label{secIV}

We have performed a comparative study of the charmed baryon$-$nucleon interaction 
based on different theoretical approaches.
For this purpose, we make use of i) a constituent quark model tuned in 
the light-flavor baryon$-$baryon interaction and the hadron spectra,
ii) hadronic descriptions based on one-boson exchange potentials, 
iii) a quark delocalization color screening model,
iv) (2+1)-flavor lattice QCD results
of the HAL QCD Collaboration at unphysical pion masses 
and their effective field theory extrapolation to the physical pion mass.
There is a general qualitative agreement among the different 
available approaches to the charmed baryon$-$nucleon interaction. 
Quark-model based results point to soft interactions without two-body bound states. They
also support a negligible channel coupling, due either to tensor forces
or transitions between different physical channels, $\Lambda_c N - \Sigma_c N$.
The short-range dynamics of the CQM model, fixing the hard-core
radius of the $S$ wave interactions, generates a slightly larger repulsion in the $^1S_0$ 
than in the $^3S_1$ $\Lambda_c N$ partial wave.
A similar asymmetry between the attraction in the two $S$ waves of the $\Lambda_c N$ interaction also appears 
in hadronic approaches.

Pauli suppression effects generate a major impact
in the $\Sigma_c N$ charmed baryon$-$nucleon interaction, 
resulting in a strong repulsion in
the $^1S_0 (I=1/2)$ and $^3S_1 (I=3/2)$ partial waves. 
A comparative detailed study of Pauli suppressed partial waves, as the $^1S_0 (I=1/2)$ 
and $^3S_1 (I=3/2)$ $\Sigma_c N$ channels, would help to disentangle the short-range 
dynamics of two-baryon systems containing heavy flavors.
Quark-model approaches predict a small contribution of 
the channel coupling to the charmed baryon$-$nucleon interaction, 
concluding that the $\Lambda_c N$ tensor potential is negligibly weak
and that the coupling between $\Lambda_c N$ and $\Sigma_c N$ channels is also weak.

In the light of the results for the $Y_c N$ interactions, the possible existence of charmed 
hypernuclei has been discussed. The order of the isoscalar $J=1/2$ and $J=3/2$ channels
is reversed in the charm with respect to the strange sector. While the existence of an isoscalar $J=1/2$ $\Lambda_c NN$ charmed
hypernucleus is not likely, that of an isoscalar $J=3/2$ state seems more feasible. In any case,
the possible existence of $\Lambda_c$ hypernuclei in light or medium$-$heavy nuclei
is a firm prediction of quark-model and hadronic approaches
to the $Y_c N$ interaction.

The understanding of the baryon$-$baryon interaction in the heavy flavor
sector is a key ingredient in our quest to describing the properties of hadronic matter. 
The study of unknown two-baryon systems could benefit from well-constrained models 
based as much as possible on symmetry principles and analogies with other similar 
processes. Subsequently, lattice QCD simulations could incorporate firmly established predictions 
to validate our understanding of low-energy Quantum Chromodynamics in the multiquark sector.

\section{Acknowledgments}
This work has been partially funded by COFAA-IPN (M\'exico) and 
by Ministerio de Econom\'\i a, Industria y Competitividad 
and EU FEDER under Contract No. FPA2016-77177-C2-2-P.

\end{document}